\documentstyle[epsf]{mn}

\def\Gyr{{\rm\,Gyr}}

\def\kms{{\rm\,km\,s^{-1}}}
\def\kpc{{\rm\,kpc}}
\def\mpc{{\rm\,Mpc}}
\def\msun{{\rm\,M_\odot}}

\begin{document}

\title{Globular Cluster Evolution in M87 and Fundamental Plane Ellipticals}

\author[Chigurupati Murali and Martin D. Weinberg]{Chigurupati Murali
 and Martin D. Weinberg\thanks{Alfred P. Sloan Foundation Fellow.}\\
 Department of Physics and Astronomy, University of Massachusetts,
 Amherst, MA 01003-4525, USA}

\maketitle

\begin{abstract}

	The globular cluster population in M87 has decreased
measurably through dynamical evolution caused by relaxation, binary
heating and time-dependent tidal perturbation.  For fundamental plane
ellipticals in general, cluster populations evolve more rapidly in
smaller galaxies because of the higher mass density.  A simple
evolutionary model reproduces the observed trend in specific frequency
with luminosity for an initially constant relationship.

	Fits of theoretically evolved populations to M87 cluster data
from McLaughlin et al. (1994) show the following: 1) dynamical effects
drive evolution in the initial mass and space distributions and can
account for the large core in the spatial profile as well as producing
radial-dependence in the mass spectrum; 2) evolution reduces $S_N$ by
50\% within $16\kpc$ and 35\% within $50\kpc$, implying that $S_N$ was
initially 26 in this region.  We estimate that 15\% of the `missing'
clusters lie below the detection threshold with mass less than $10^5
M_{\odot}$.

\end{abstract}

\begin{keywords}
globular clusters: general -- galaxies: individual (M87) -- galaxies:
star clusters
\end{keywords}

\section{Introduction}
\label{sec:intro}

	Observations of some giant elliptical galaxies reveal globular
cluster systems which appear more extended than the host (Harris
1991).  A particularly well-documented example belongs to M87 with a
core radius of 7 arc sec and a cluster system with a core radius of 1
arc min (McLaughlin 1995).  However, because a cluster population
evolves dynamically due to both internal and external processes, the
currently observed population almost certainly differs from the
primordial one, complicating the interpretation.

	Researchers have attempted to explain the extended core of the
M87 cluster distribution as the evolved remnant of an initial profile
which more closely resembled the light.  However, neither dynamical
friction nor shocking by a compact nucleus can fully account for this
feature.  Lauer \& Kormendy (1986) found that a dynamical friction
induced inflow can broaden an initially peaked spatial distribution
but not at the observed scales.  Ostriker, Binney \& Saha (1989,
hereafter OBS) subsequently determined that nuclear tidal disruption
is viable only if clusters formed exclusively on box orbits.

	Another potential mechanism is cluster evaporation through
dynamical evolution.  Recent work in this area demonstrates that
evaporative mass loss driven by relaxation and heating due to a
time-varying tidal field can lead to strong evolution of the Milky Way
cluster population in a Hubble time (Weinberg 1994, Murali \& Weinberg
1996, hereafter MW).  In this paper, we examine these influences on
cluster evolution in the dense inner regions of M87 and find that they
produce the observed flattened profile from a peaked initial
distribution over a wide range of initial conditions.  Direct
estimates of initial conditions using dynamically evolved parametric
models of the spatial distribution and cluster mass function indicate
that roughly 35\% of the initial population dissolves or evolves below
the detection threshold leaving the large core as a result.
Furthermore, the decay in the size of the cluster population
corresponds to a decrease in the specific frequency of globular
clusters, $S_N$, which denotes the number of clusters per unit
galaxian luminosity with $L$ measured in units of $M_v=-15$.

	The high values of $S_N$ found in giant ellipticals have
become a key point in galaxy formation arguments and suggest, for
example, that the cluster system formed along with M87 (e.g. Harris
1991; van den Bergh 1995).  Here we show that cluster systems decay
more rapidly in less luminous fundamental plane ellipticals; this
leaves larger values of $S_N$ in luminous galaxies at the present
epoch even if all ellipticals begin with equal $S_N$.  Our results
thus provide at least a partial explanation for the observed trend of
$S_N$ with $L$.

The plan of the paper is as follows.  We summarize our choices for the
cluster population and the mass model for M87 in \S\ref{sec:pop}.  The
assumptions and method for dynamically evolving the population is
presented in \S\ref{sec:evol}.  The main results, the statistical
comparison of the observed clusters to the theoretical models, is
presented in \S\ref{sec:results}.  This includes an exploration of the
evolutionary trends, best fit spatial profiles and mass functions, and
an inference of the primordial population.  In \S\ref{sec:disc}, we
discuss the discuss the importance of the fundamental plane properties
on the observed relation between specific frequency and luminosity
A summary is given in \S\ref{sec:summary}.

\section{Cluster population}
\label{sec:pop}

	We assume that the cluster population formed in an initial
burst approximately $11 \Gyr$ ago.  Stellar evolution dominated
cluster evolution for the first $\Gyr$ for a Salpeter IMF
($\beta=2.35$) with $m_l=0.1 \msun$, corresponding to the main
sequence lifetime of $2\msun$ A-star.  Our zero-age population
represents the epoch when, approximately $10\Gyr$ ago, relaxation,
external heating and core collapse heating began to drive cluster
evolution.  

	The fiducial calculations represent zero-age clusters as
$W_0=5$ King models.  Comparison calculations using $W_0=7$ clusters
show nearly identical evolution over the long time scales of interest,
in agreement with the results of MW where overall evaporation times
were found to depend weakly on concentration in the range $5\leq
W_0\leq 7$.  We expect similar trends in evolution for $W_0=3$
clusters (c.f. MW), except in high mass, low-eccentricity cases where
tidal heating leads to rapid disruption.  These clusters enhance the
destruction rate described below, but constitute a very small fraction
of expected initial populations.

	Each cluster is tidally limited on its orbit in the host.
While initial cluster densities may differ from the mean density
required by perigalactic tidal limitation, subsequent evolution during
the first $\Gyr$ leads rapidly to tidal truncation or disruption.  The
limiting or {\it tidal} radius $R_T$ is uniquely determined by the
cluster mass and orbit.  Table \ref{tab:init} summarizes the choice of
parameters for individual clusters.

\begin{table*}
\caption{Cluster Initial Conditions}
\label{tab:init}
\begin{tabular}{lllc} 
\multispan3 Structural parameters\hfill&Fiducial value\hfill\\ \hline
$M$&& total mass \hfill&$10^5\leq M_c\leq5\times 10^6 M_{\odot}$ \\
$W_0$&& King concentration parameter \hfill&$W_0=5, 7$\hfill \\
$R_T$&& cluster limiting radius \hfill&tidal limitation\hfill \\ \hline
\multispan3 Mass spectral parameters\hfill&\\ \hline
$\beta$&& mass spectral index: $N(m)\propto m^{-\beta}$ \hfill&
	$\beta=2.35$ (Salpeter)\hfill\\
$m_l$&& lower mass limit \hfill&$m_l=0.1$\hfill\\
$m_u$&& upper mass limit \hfill&$m_u=2.0$\hfill\\ \hline
\multispan3 Orbital parameters\hfill\\ \hline
E&& orbital energy\hfill&isotropic orbit\hfill\\
$\kappa$&& relative ang. mom.: $J/J_{max}(E)$\hfill&distribution\hfil\\
&&\hfill&\\
\end{tabular}
\end{table*}

To represent the cluster mass distribution, $\nu(M,r)$, we use pure
power laws (e.g. Harris \& Pudritz 1994), power laws whose exponents
have a linear dependence on radius, and a Gaussian magnitude
distribution (e.g. McLaughlin, Harris \& Hanes 1994, MHH).  Power law
mass distributions have been proposed on physical grounds by Harris \&
Pudritz (1994) while the Gaussian is commonly used as a convenient
fitting function for the observed cluster luminosity function.  To
represent the spatial distribution of the cluster population in the
primary, we use power law densities with and without a core derived
from isotropic distribution functions, $f(E)$.  Orbital isotropy is
assumed due to lack of observational constraint.

	Adopted models are given by joint distributions
$\nu(M,r)\times f(E)$ and are summarized in Table \ref{tab:pop}.  The
Model 1 and Model 2 families use power law mass and Gaussian magnitude
distributions respectively.  Within each family, successive models
have additional parameters to explore varying core sizes and radial
dependence of the mass spectral index.  Detailed derivation of models
from the underlying distribution function is given in Appendix
\ref{sec:modAp}.

\begin{table*}
\caption{Population models}
\label{tab:pop}
\begin{tabular}{lccl}
Designation&$\rho(r)$&$\nu(M)$&Parameters\\
\hline
Model 1a\hfill&\hfil$\rho_0 r^{-\eta}\hfil$&\hfil$M^{-\alpha}$\hfil&
	$\eta,\alpha$\hfill\\
Model 1b\hfill&\hfil$\rho_0 r^{-\eta}\hfil$&\hfil$M^{-(\alpha+Kr)}$\hfil&
	$\eta,\alpha,K$\hfill\\
Model 1c\hfill&\hfil$\rho_0(r_c^2+r^2)^{-\eta/2}$
	\hfil&\hfil$M^{-\alpha}$\hfil&$\eta,r_c,\alpha$\hfill\\
Model 1d\hfill&\hfil$\rho_0(r_c^2+r^2)^{-\eta/2}$
	\hfil&\hfil$M^{-(\alpha+Kr)}\hfil$\hfil&$\eta,r_c,\alpha,K$\hfill\\
\hline
Model 2a\hfill&\hfil$\rho_0 r^{-\eta}\hfil$&\hfil$e^{-(V-V_0)^2/2\sigma_V^2}
	\cdot dV/dM$\hfil&$\eta,V_0,\sigma_V$\hfill\\
Model 2b\hfill&\hfil$\rho_0(r_c^2+r^2)^{-\eta/2}$
	\hfil&\hfil$e^{-(V-V_0)^2/2\sigma_V^2}\cdot dV/dM$\hfil&
	$\eta,r_c,V_0,\sigma_V$\hfill\\
\end{tabular}
\end{table*}

Finally, we represent the potential of M87 as a singular isothermal
sphere, with rotation velocity $v_0=606\kms$ (e.g. OBS), velocity
dispersion $\sigma=350\kms$, and assume a distance of 16 Mpc (van der
Marel 1992).  This defines a length scale of 77.6 pc per second of
arc.  Further discussion of potential and distance scale is given in
Appendix \ref{sec:isoAp}.

\section{Cluster evolution}
\label{sec:evol}

Competition between internal relaxation and heating due to external
forcing may dramatically affect a cluster's evolutionary time scale
and survival history.  In addition to impulsive heating of a cluster
halo---in a gravitational bulge shock, for example---resonances
between the cluster's own orbital motion and internal stellar
trajectories may heat cluster stars beyond the limit set by adiabatic
invariance (Weinberg 1994).  For tidally-limited clusters resonant
heating on low-eccentricity orbits and tidal limitation on
high-eccentricity orbits drive rapid cluster evolution and evaporation
(see MW for details).  The strength of these effects motivates this
study.

The evolution of individual clusters includes two-body relaxation in
the one-dimensional Fokker-Planck approximation (e.g. Cohn 1979),
external heating due to the time-varying tidal field (MW), and a
phenomenological binary heating term (e.g. Lee et al.  1991).

We take advantage of the scale-free galaxian profile by fixing orbital
energy $E$ of all clusters, choosing an initial grid of
tidally-limited clusters in $\kappa=J/J_{max}(E)$ and mass, and
computing the evolution to complete evaporation.  The quantity
$J_{max}(E)$ denotes the maximum angular momentum of an orbit with
energy $E$.  This grid may then be scaled to all desired orbital
energies.  The time evolution of the space density for the entire
population is then constructed by determining the phase space
distribution at the desired time using the evolutionary calculations
and projecting appropriately.

Although we specifically consider M87, the results apply to any
elliptical with similar profile.  For example, we can scale evolution
to any fundamental plane elliptical.  Because the period decreases
with mass, the same initial population will be more evolved for
smaller mass primaries (see \S\ref{sec:disc} for more discussion).

\section{Results}
\label{sec:results}

\subsection{Evolution of the core}
\label{sec:core}

\begin{figure}
\epsfxsize=20pc
\epsfbox[12 138 600 726]{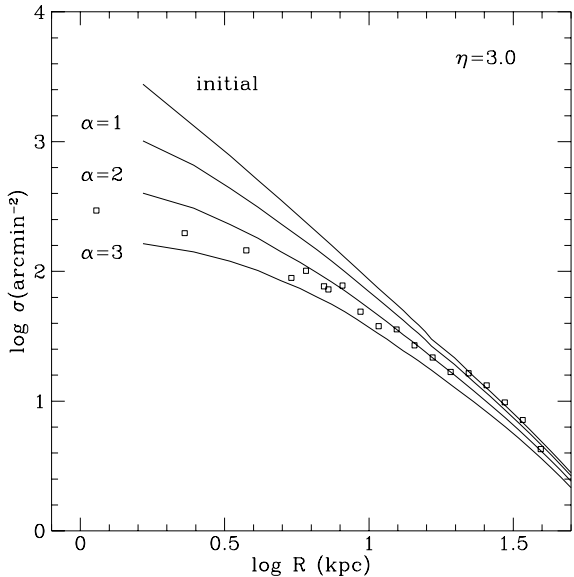}
\caption{
	Surface density inferred from observations (open squares)
	compared to Model 1a for indicated values of mass spectral
	index $\alpha$ and initial $r^{-3}$ ($\eta=3$) profiles (solid
	curves).  Rapid evolutionary rates of low mass clusters
	produce flatter cores for larger $\alpha$.
}
\label{fig:e3.comp}
\end{figure}

	An initially peaked cluster distribution develops a flattened
core through dynamical evolution of individual clusters, as shown in
Figure \ref{fig:e3.comp}.  In the dense regions of the inner galaxy,
rapid mass loss due to relaxation and tidal heating can cause complete
evaporation of a cluster or drive it below the observational limit.
Rapid relaxation results from the high densities imposed by tidal
limitation while tidal heating strongly enhances evaporation rates on
low-eccentricity orbits.  The resulting profiles are similar to the
profile derived in McLaughlin (1995).

The overall shape of the evolving profile depends on both the initial
mass and space distribution of the clusters.  Consider the following
limits.  For a fixed spatial profile, a distribution rich in low mass
clusters evolves rapidly due to short evaporation times while for a
fixed mass distribution, a sharply peaked spatial profile develops a
smaller core than a shallow profile.  Taken together, the two trends
produce a correlation between the inferred initial mass distribution
and density profile: a large population of low mass clusters can
rapidly flatten a steep initial profile while, conversely, a large
population of high mass clusters allows a flatter initial profile to
evolve slowly to the same final shape.  Observations of the cluster
mass distribution will distinguish between the different initial
conditions.

\subsection{Estimates of initial conditions}
\label{sec:param}

	The evolved cluster populations described in \S\ref{sec:evol}
are compared to observed cluster data\footnote{The data have been
kindly supplied by McLaughlin \& Harris (1995)} using a maximum
likelihood estimator which combines model and background surface
densities with incompleteness measurements.  The background surface
density is taken to be 6.33 per arcmin$^2$ with a uniform luminosity
function (MHH).  Point sources lie in the region $1.21'\leq R \leq 7'$
(the field edge), centered on M87, with apparent limiting magnitude
$V=24$.  We use a mass-to-light ratio $(M/L)_V=2$ to convert
luminosity to mass.  Note that larger $M/L$ will shift the population
to higher mass, implying less evolution, while smaller $M/L$ will have
the opposite effect.  For $R<1.21'$, the authors provide 3 binned
points (McLaughlin 1995).  A joint $\chi^2$-likelihood estimator is
used to include all data points.  Details of the estimation procedure
are provided in Appendix \ref{sec:mlAp}.

	We fit both dynamically evolved distributions and unevolved
distributions based on the models presented in Table \ref{tab:pop}.
In the case of dynamically evolved models, the quoted values represent
initial conditions (labeled by `initial').  In the case of unevolved
models, the quoted values represent the best fit parameters at the
present epoch (labeled by `present').  Only models with cores are
considered in the present epoch fits because coreless models poorly
represent the data.

	Tables \ref{tab:mod1fits} and \ref{tab:mod2fits} present the
best estimates and their variances (cf. Table \ref{tab:pop}).
Comparison of present epoch and initial parameter estimates
illustrates several expected evolutionary trends.  The core of the
distribution grows due to the depletion of clusters in the inner
regions of the galaxy.  The power law index $\alpha$ decreases, while
the Gaussian magnitude peak $V_0$ and slope $K$ of the
radially-dependent power law both increase as a result of the
selective evaporation of lower mass clusters.  The increase in $K$
also indicates that depletion occurs primarily in the inner regions.
The following cases show specific features of these trends.

\begin{table*}
\caption{Model 1 fits}
\label{tab:mod1fits}
\begin{tabular}{lccccccccc}
Epoch&$\eta$&$\sigma_{\eta}$&$r_c$&$\sigma_{r_c}$&$\alpha_0$&$\sigma_{\alpha_0}$&
$K$&$\sigma_{K}$&-log L\\
\hline
\multispan{10}\hfil Model 1a \hfil\\
\hline
initial&2.76&0.04&-&-&1.95&0.03&-&-&69502.9\\
\hline
\multispan{10}\hfil Model 1b\hfil\\
\hline
initial&2.66&0.06&-&-&1.68&0.09&0.01&0.004&69501.3\\
\hline
\multispan{10}\hfil Model 1c\hfil\\
\hline
present&3.09&0.13&7.30&0.79&1.73&0.04&-&-&69500.6\\
initial&3.03&0.12&5.67&0.94&1.93&0.04&-&-&69499.8\\
\hline
\multispan{10}\hfil Model 1d \hfil\\
\hline
present&3.06&0.13&7.34&0.81&1.30&0.08&0.028&0.002&69480.9\\
initial&3.13&0.10&5.70&0.40&1.61&0.10&0.014&0.004&69492.8
\end{tabular}
\end{table*}

\begin{table*}
\caption{Model 2 fits}
\label{tab:mod2fits}
\begin{tabular}{lccccccccc}
Epoch&$\eta$&$\sigma_{\eta}$&$r_c$&$\sigma_{r_c}$&$V_0$&$\sigma_{V_0}$&
	$\sigma_V$&$\sigma_{\sigma_V}$&-log L\\
\hline
\multispan{10}\hfil Model 2a\hfil \\
\hline
initial&2.77&0.05&-&-&-7.11&0.17&1.16&0.08&69487.2\\
\hline
\multispan{10}\hfil Model 2b\hfil \\
\hline
present&3.10&0.13&7.32&0.80&-7.33&0.10&1.08&0.07&69468.7\\
initial&3.14&0.12&5.14&0.77&-7.07&0.17&1.19&0.08&69485.3
\end{tabular}
\end{table*}

\begin{figure}
\epsfxsize=20pc
\epsfbox[12 138 600 726]{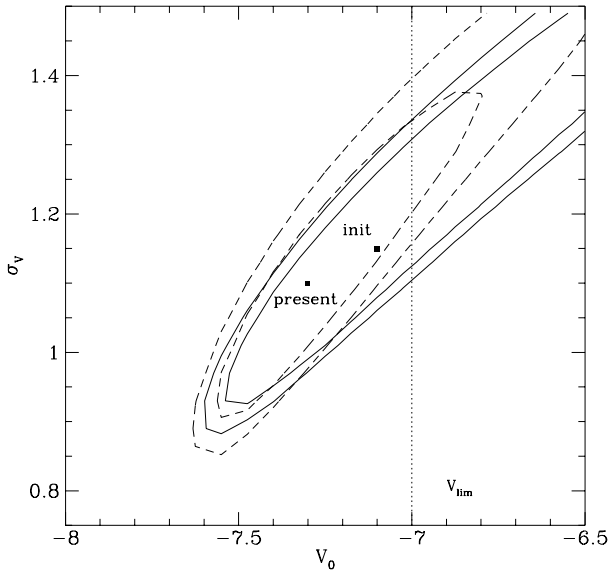}
\caption{
	95\% and 99\% confidences for marginal density in $V_0$ and
	$\sigma_V$ for present (dashed) and initial (solid) Model 2a
	fits.  Points indicate best-fit values for present and initial
	models.  The magnitude limit of the data is indicated as the
	vertical line.
}
\label{fig:g.comp}
\end{figure}

	We plot the marginal probability density in $V_0$ and
$\sigma_V$ for Model 2b in Figure \ref{fig:g.comp}.  The best estimate
for $V_0$ decreases with time as low mass (and therefore high $V$)
clusters disappear.  However, the assumption of identical initial and
present epoch values cannot be ruled out since the values of $V_0$ are
only weakly inconsistent (cf. Table \ref{tab:mod2fits}).  The lack of
constraint could result from the shallow magnitude limit of the data.
Both fits are consistent with distributions which peak below the
limiting magnitude of $V=-7$.

\begin{figure}
\epsfxsize=20pc
\epsfbox[12 138 600 726]{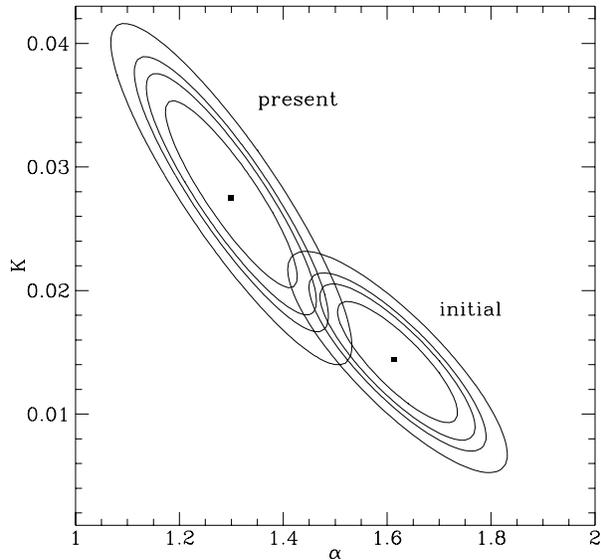}
\caption{75\%, 90\%, 95\%, and 99\% confidences for marginal density 
		in $\alpha$ and $K$ for present epoch and initial
		Model 1d fits.  Points mark best-fit values.  The
		initial spectral index shows mild radial dependence
		increasing from a central value of 1.61 to 1.97 at 30
		kpc in the best-fit case.
}
\label{fig:cih.comp}
\end{figure}

	Model 1d suggests that there is radial dependence in the mass
distribution (Fig. \ref{fig:cih.comp}).  Both present epoch and
initial fits are inconsistent with a constant mass spectral index
($K=0$) and indicate that dynamical evolution has enhanced the radial
dependence.  In the core region, the present index $\alpha\approx
1.4$.  We note that these results conflict with those of MHH and
McLaughlin \& Pudritz (1996), who find no radial dependence in the
mass distribution.

\subsection{Comparison of models}
\label{sec:comp}

The previous section examined the results of trends in the evolution
of the cluster population within several model families.  Here, we
identify the best overall representation among the initial models
after $10\Gyr$ of evolution using a generalized likelihood ratio test
(e.g. Martin 1971).

The final column in Table \ref{tab:mod1fits} shows that the most
general model gives a better estimate in each case, as expected.  In
particular, Model 1c can be rejected in favor Model 1d, a result which
is consistent with the confidence surfaces for the radially-dependent
mass spectrum plotted in Figure \ref{fig:cih.comp}.

\begin{figure}
\epsfxsize=20pc
\epsfbox[12 138 600 726]{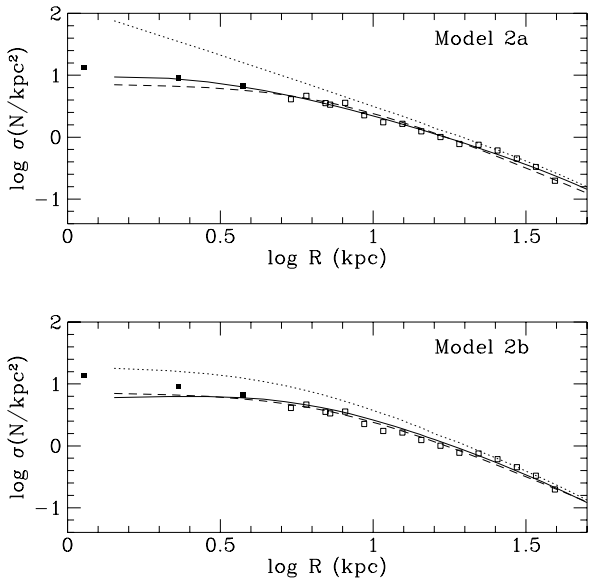}
\caption{Comparison of surface density profiles of Models 2a and 2b (solid) 
		with binned data.  Initial profiles (dotted) and
		present epoch Model 2b fits (dashed) are also plotted.
		Both the model with initial core and the present epoch
		fit deviate from the data in the inner region.
}
\label{fig:model2.comp}
\end{figure}

Model 2b generalizes Model 2a by introducing arbitrary $initial$ core
size.  A finite core does provide a better estimate but zero core (Model
2a) cannot be rejected.  Figure \ref{fig:model2.comp} compares the
surface density profiles of these two models.  The Model 2b fit falls
below the observed profile at small radii due to the shallow initial
core.  However, the binned surface density points have relatively low
weight in the full data set.  The good fit of Model 2a to the inner
data points suggests that a more peaked initial distribution may
provide the optimal fit.  The use of individual cluster counts in this
region should help provide the necessary constraint.  The deviation of
the present epoch fit from the data further suggests that evolution
plays an important role in shaping the profile.

Finally, we compare the most general power law mass function model,
Model 1d, with the most general Gaussian magnitude model, Model 2b, by
constructing a linear combination of both spaces and searching for the
global maximum.  The maximum occurs at the best-fit parameters for
Model 2b: the Gaussian magnitude distribution describes the data
significantly better than any power law mass distribution.  As
discussed above, the Gaussian may be poorly constrained by the
limiting magnitude of the data.  However, examination of the estimated
functions shows that the power law is more peaked at low mass than the
Gaussian both initially and finally, while the Gaussian has more
weight at high mass.  Thes differences in shape also lead to the
statistical preference of the Gaussian luminosity function over the
spatially constant power law and its radially-dependent
generalization.

Table \ref{tab:ratio} summarizes the conclusions of the tests
comparing the initial conditions and lists the likelihood ratios and
confidence values.  In the first two cases, the likelihood ratios
follow directly from Tables \ref{tab:mod1fits} and
\ref{tab:mod2fits}. In summary, we find that the initial Gaussian
magnitude distribution best describes the data, but that we cannot
distinguish between singular density profiles and densities with core.
Both conclusions may result from insufficient data.

\begin{table}
\caption{Evolved model comparisons}
\label{tab:ratio}
\begin{tabular}{lccccc}
test&$-2\ln \lambda^\flat$&$\nu^\sharp$&accept&reject&confidence\\
\hline 
1c--1d &14.0&4&&$\surd$&99\%\\
2a--2b &3.8&4&$\surd$&&60\%\\
1d--2b &15.0&7&&$\surd$&96\%\\ 
\hline
\multispan{6} {\tiny $^\flat-2\ln\lambda$ is likelihood ratio}\hfill\\
\multispan{6} {\tiny $^\sharp\nu$ is number of degrees of freedom}\hfill
\end{tabular}
\end{table}

\subsection{Evolution of the initial population}
\label{sec:evolpop}

\begin{figure}
\epsfxsize=20pc
\epsfbox[12 138 600 726]{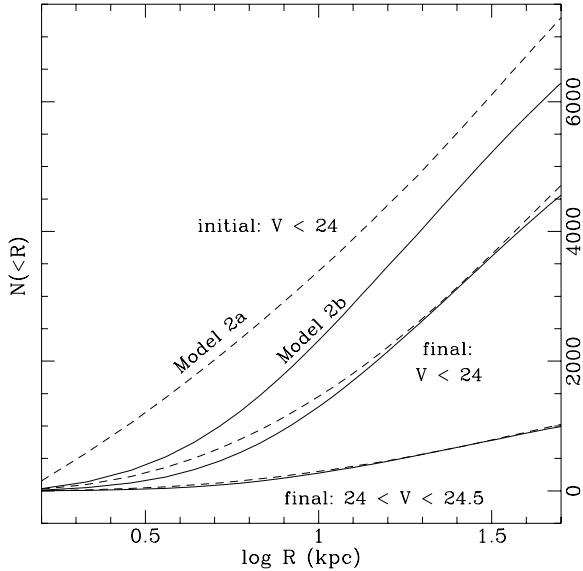}
\caption{Estimated cumulative distribution of clusters for projected
		radius $R<50 \kpc$ $(6.5R_e)$.  Initial distributions
		for Model 2a (dashed) and Model 2b (solid) are shown
		for $V<24$ (upper pair).  Final distributions are
		shown for $V<24$ (middle pair) and for $24>V>24.5$
		(bottom pair).  Clusters with $24>V>24.5$ began with
		$V<24$.  Approximately 50\% of the initial clusters
		vanish within (16 kpc) $2R_e$ and 35\% within $6.5R_e$
		for Model 2a.}
\label{fig:N.comp}
\end{figure}

	From the derived initial conditions, we plot the projected
cumulative distribution of clusters initially within 50 kpc or
$6.5R_e$ (Figure \ref{fig:N.comp}). The initial population in this
region is about 7250 for Model 2a, a factor of 1.6 larger than the
presently observed population of 4500. For Model 2b, the initial
population size is smaller, but still in excess of 6000.  Evolved
clusters will also be found at masses below the observational limit.
Within 16 kpc, about 500 evolved clusters are expected in the range
$24>V>24.5$ ($10^5 > M > 7\times 10^4 M_{\odot}$).  All other clusters
in this region initially in this mass range will have evaporated.

\begin{figure}
\epsfxsize=20pc
\epsfbox[12 138 600 726]{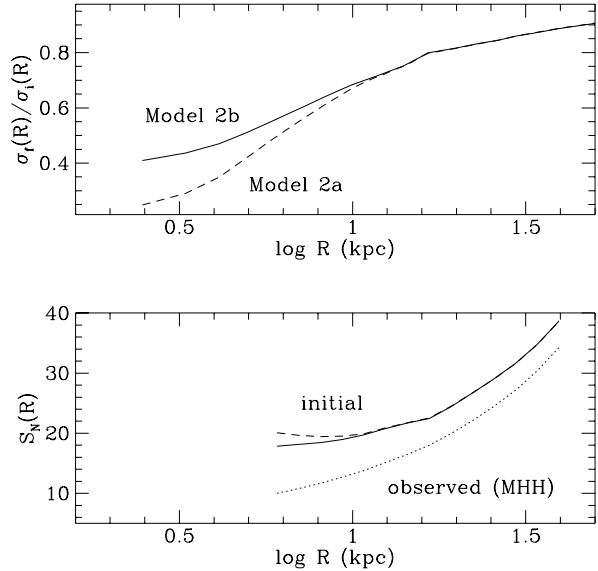}
\caption{The ratio of final to initial surface density (top) in 
	Models 2a (dashed) and 2b (solid) and the run in initial
	specific frequency at $R$ ($S_N(R)$) for each model (bottom)
	derived from the observed values given by MHH (dotted).
	Evolution reduces $S_N$ by 35\% in Model 2a.}
\label{fig:sn.comp}
\end{figure}

	This implies that the specific frequency, $S_N$, evolves with
time.  Using the ratio of final to initial surface densities from the
models, we derive the run in initial specific frequency at radius $R$,
$S_N(R)$ (Figure \ref{fig:sn.comp}).  As expected, depletion in the
inner regions dominates; however, $\sim10\%$ change in $S_N(R)$ occurs
even out to 50 kpc due to the rapid evolution of low-mass clusters.

\begin{figure}
\epsfxsize=20pc
\epsfbox[12 138 600 726]{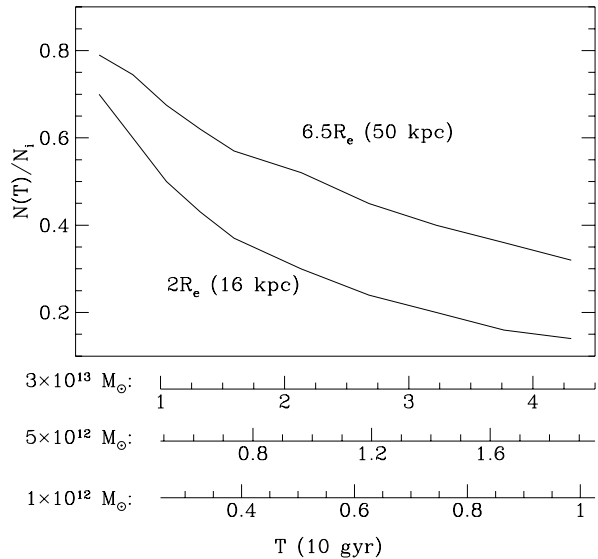}
\caption{The fraction of clusters remaining within the indicated projected
	radius as a function of time.  Abscissae indicate time units
	which correspond to fundamental plane scaling for galaxies
	with indicated masses.  Top axis gives M87 scaling (other axes
	will be discussed in the next section).  Initial $S_N$ is
	$\approx 21$ measured within 16 kpc and 26.5 measured within
	50 kpc and decays with the cluster population.}
\label{fig:N_t}
\end{figure}

	Model 2a in Figure \ref{fig:N.comp} shows a 35\% change in
total $S_N$ within 50 kpc due to depletion.  The observed total value
$S_N\approx17$ in this region (MHH) implies an initial value of 26.5.
Thus even the enormous observed value of $S_N$ has diminished
significantly due to evolution (this neglects the intrinsic evolution
in galaxy luminosity due to stellar evolution).  The time evolution of
the total $S_N$ is shown directly in Figure \ref{fig:N_t}.  The decay
of the cluster population is approximately exponential in time with
e-folding times of $20 \Gyr$ and $40 \Gyr$ for measurements within 16
and 50 kpc, respectively.

	Our comparison applies to clusters in specified mass and
radial ranges in a galaxy.  Quoted specific frequency values are
extrapolated over unobserved ranges from data taken within such
limits.  In the case of M87, MHH find $S_N=17.7$ directly from the
observations, while extrapolation yields a total $S_N=14.4$ for the
whole galaxy.  Their extrapolation of the luminosity function over all
magnitudes yields a total correction of 2.2 to cluster counts in the
observed magnitude range.  For $R<42 \kpc$, using this correction they
estimate a total $\sim9400$ clusters (3729 observed in galaxy+cD
envelope, 500 in core, $\times 2.2$) which is 70\% of the total number
of clusters estimated over all radii.  The estimated initial
distribution has a larger correction because it is weighted more
heavily to low mass (c.f. \S\ref{sec:param}).  Since most of these
clusters completely evaporate, this implies even greater evolution in
specific frequency than derived here.

\section{Discussion}
\label{sec:disc}
	
	Our conclusions ignore the possibility that recent merger and
accretion events have have strongly contaminated the initial cluster
population.  Merging of gas-rich galaxies is expected to produce
clusters with strong central concentration (Zepf \& Ashman 1993; Mihos
\& Hernquist 1994), but the large core of the distribution itself
argues against any recent merger which has produced significant
numbers of young clusters.  Thus either cluster-producing mergers in
M87 have occurred long in the past or not at all.

The recent addition of clusters through satellite accretion is also
unlikely to account for a significant fraction of the observed
population.  For example, accretion of Milky Way-type spirals can only
account for about 25\% of the observed population, assuming that the
$4 L_*$ luminosity of M87 ($R<40 \kpc$) comes entirely from satellites
(Lauer 1988).  Assuming that material is stripped when the mean
density of the primary exceeds that in the satellite, accretion will
deposit clusters in regions of mean density similar to that in the
original environment of the accreted galaxy.  The clusters, then, will
remain roughly tidally truncated.  Their new orbits depend on the
orbit of the dissolving satellite, but otherwise evolution should be
similar and the accreted population may appear coeval regardless of
the time of accretion.

	The high specific frequency of globular clusters in M87
appears to indicate the exceptional conditions governing the formation
and evolution of cD galaxies relative to other ellipticals.  The
observation that specific frequency increases with galaxy luminosity
suggests that galaxy formation was not an intrinsically hierarchical
and homologous process (e.g. Santiago \& Djorgovski 1993).

	However, the current results indicate that density differences
between galaxies will lead to differences in the evolution of
intrinsic cluster populations.  Since M87 lies approximately on the
fundamental plane for elliptical galaxies, we can investigate
differences in environment-driven cluster evolution by scaling our
results to other ellipticals.  We assume that the initial profile of
the cluster population derived above, when scaled homologously,
describes the initial population in any elliptical.  The dynamical
time scale for a tidally truncated cluster with constant mass and
spatial profile is then determined by the mean density of the galaxy:
$\tau\sim M^{-1/2}R^{3/2}$.  This yields $\tau\sim M^{\beta}$, where
$\beta=0.4$ for the fundamental plane scaling given in Faber et al.
(1987), $\beta=0.26$ for the Djorgovski \& Davis (1987) results, and
$\beta=0.6$ for a more recent set of parameters from Faber (1995; see
also Pahre et al. 1995).  This relation implies that clusters evolve
and are depleted more rapidly in smaller, low luminosity ellipticals
than in massive, high luminosity ellipticals.

	To demonstrate the importance of this effect, we combine the
approximate exponential decay rate of the cluster population found in
\S\ref{sec:evolpop} with this scaling relation.  This gives an
expression for the number of clusters remaining in an initially coeval
population belonging to a galaxy of luminosity $L$ at time $T$:

\begin{eqnarray}
N_{cl}(L,T)=N_0(L)e^{-T/\tau_0(M_{M87}/M)^{\beta}}\nonumber\\
        =N_0e^{-T/\tau_0(L_{M87}/L)^{1.24\beta}},
\end{eqnarray}
where $\tau_0\approx 40 \Gyr$ for M87, $N_0(L)$ is the initial
distribution of cluster population sizes as a function of galaxy
luminosity and $M/L\sim L^{1.24}$ throughout.  

\begin{figure}
\epsfxsize=20pc
\epsfbox[12 138 600 726]{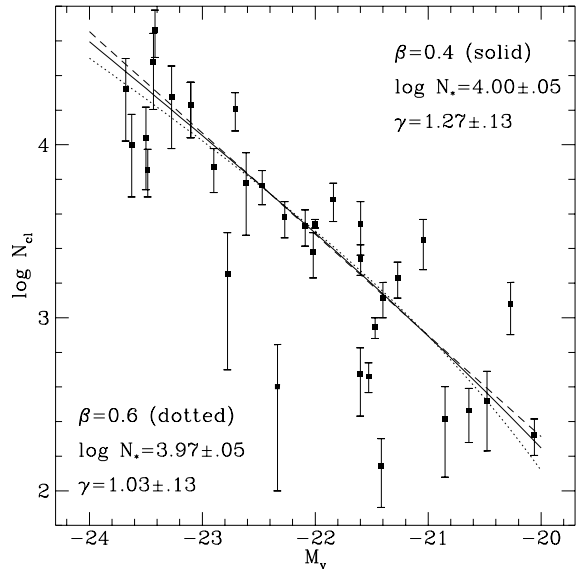}
\caption{Evolved population size as a function of galaxy luminosity 
	compared to data from Harris (1996) for fundamental plane
	parameter $\beta=0.4$ (solid) and $\beta=0.65$ (dotted) .  The
	present-day, unevolved fit (dashed) has $\log N_*=3.89\pm.05,
	\gamma=1.46\pm.13$.  For $\beta=0.65$, the initial value of
	$\gamma$ is consistent with specific frequency which is
	independent of luminosity}
\label{fig:fpe.1}
\end{figure}

	Using a power-law distribution
$N_0(L)=N_{*}(L/L_{M87})^\gamma$, we compare the model curve
$N_{cl}(L,T)$ to an observed sample of cluster populations in galaxies
(Harris 1996). The resulting models show qualitative agreement with the
data (Figure \ref{fig:fpe.1}), falling off at low luminosity more
rapidly with increasing $\beta$ due to the more rapid rate of
evolution.  From this we conclude that the observed discrepancies in
specific frequency which correlate with galaxy luminosity were smaller
in the past.  This is reflected in the smaller exponent $\gamma$ in
$N_{cl}\propto L^{\gamma}$ predicted initially compared to the
present-day estimate.

	To conclude, we suggest that cluster evolution may account at
least in part for the specific frequency problem.  Because clusters
evolve more rapidly in dense environments and because ellipticals are
typically denser at low luminosity that at high luminosity, cluster
populations diminish more rapidly in low luminosity galaxies.
Specific frequencies will therefore correlate more strongly with
luminosity at later times.

\section{Summary}
\label{sec:summary}

	We have investigated the evolution of the M87 globular cluster
system.  The results of this study have enabled us to examine the
broader question of population evolution in fundamental plane
elliptical galaxies.  Our main conclusions follow:

\begin{enumerate}

	\item The loss of globular clusters through relaxation and
	tidally-induced evaporation accounts for the large core and
	shallow profile in cluster number distribution compared to the
	light distribution in M87.

	\item Evolution produces a radial dependence in the
	present-day mass spectrum of clusters such that higher mass
	clusters predominate in the inner regions.  The models also
	indicate an initial radial dependence in the mass spectrum.

	\item Likelihood ratio tests reject an initial power law in
	favor of an initial Gaussian but do not rule out the
	possibility of an initial core in the profile.

	\item The best-fit model for M87 has an initial population of
	$7.25\times10^3$ clusters with projected radius $R<50\kpc$,
	about 60\% more than is currently observed.  Roughly 14\% of
	the initial population are now objects of slightly less than
	$10^5 M_{\odot}$; dynamical evolution can strongly modify the
	specific frequency of globular clusters.

	\item Scaling the calculations to fundamental plane elliptical
	galaxies indicates that cluster evolution in the differing
	environments qualitatively accounts for the trend in observed
	population number versus galaxy luminosity.  Smaller galaxies
	tend to have high densities and thus more rapid evolutionary
	time scales, so their cluster populations tend to diminish
	more rapidly.

\end{enumerate}
	
Some of these inferences, especially (iii), may be biased by the lack
of detailed data in the inner galaxy.  Point source data for $r<1'$
will lead to stronger constraints on the size of the initial core and
a deeper survey will pick up low luminosity objects and further
constrain the luminosity function.

\section*{Acknowledgements}
We thank Scott Tremaine for discussion and Bill Harris, Dean
McLaughlin and Steve Zepf for providing data. This work was supported
in part by NASA award NAGW-2224.

\appendix

\section{Cluster distribution functions}
\label{sec:modAp}

	The initial cluster population is represented by joint
distributions of the phase space density, $f(E)$ and the mass spectrum
$\nu(M,r)$:
\begin{equation}
 \psi(M,r,E,J)={\partial N\over{\partial M \partial E \partial J}}
		\propto f(E,r)\nu(M,r),
\end{equation}
where $\psi(M,r,E,J)$ is the number of stars per unit mass per unit
energy per unit angular momentum at a fixed point in space.

	For the initial orbit distribution in coreless models, we
generalize the isothermal distribution employed by OBS
\begin{equation}
 f(E)={\rho_0\over (2\pi\sigma_0^2)^{3/2}}e^{-E/\sigma_0^2}, \label{eq:A2}
\end{equation}
with $\sigma_0^2=\eta^{-1} v_0^2$.  Note that $\eta^{-1}={1\over 2}$ for
the isothermal sphere and $\eta^{-1}={1\over 3}$ for the fiducial
model employed by OBS.  The background gravitational potential of M87
is taken to be a singular isothermal sphere, $\Phi(r)=v_0^2\ln r$,
independent of the cluster distribution.  The space number density of
clusters for equation (\ref{eq:A2}) is then
\begin{equation}
n(r)=\rho_0 r^{-\eta}.
\end{equation}
We generalize this profile to include a core:
\begin{equation}
n(r)=\rho_0(r_c^2+r^2)^{-\eta/2}
\end{equation}
where $r_c$ is the core radius of the system.  The isotropic cluster
distribution function follows from integral inversion (e.g. Binney \&
Tremaine 1987).

	We consider three initial distributions of cluster masses: 1)
a Gaussian distribution of initial magnitudes, $V$, which is
everywhere constant in space (e.g MHH); 2) a power law distribution of
mass, $M$, which is everywhere constant in space (e.g. Harris \&
Pudritz 1994); and 3) a power law distribution of mass with a
radially-dependent spectral index.

	The Gaussian distribution of initial magnitudes $V$ defines the
mass spectrum
\begin{equation}
\nu(M)\propto e^{-{1\over 2}({V-V_0\over \sigma_V})^2}{dV\over dM},
\end{equation}
where the transformation to mass is effected by the Jacobian, $dV/dM$.
This distribution is characterized by two parameters: $V_0$ and
$\sigma_V$, the mean and dispersion of magnitudes.  

	The simple power law mass distribution
\begin{equation}
\nu(M)\propto M^{-\alpha},
\end{equation}
depends only on the mass spectral index, $\alpha$.  We define the
following radially dependent distribution
\begin{equation}
\nu(M,r)\propto M^{-(\alpha+Kr)}
\end{equation}
whose spectral index has the central value $\alpha$ and varies
linearly with radius.

\section{Generalized isothermal sphere}
\label{sec:isoAp}

\begin{figure}
\epsfxsize=20pc
\epsfbox[12 138 600 726]{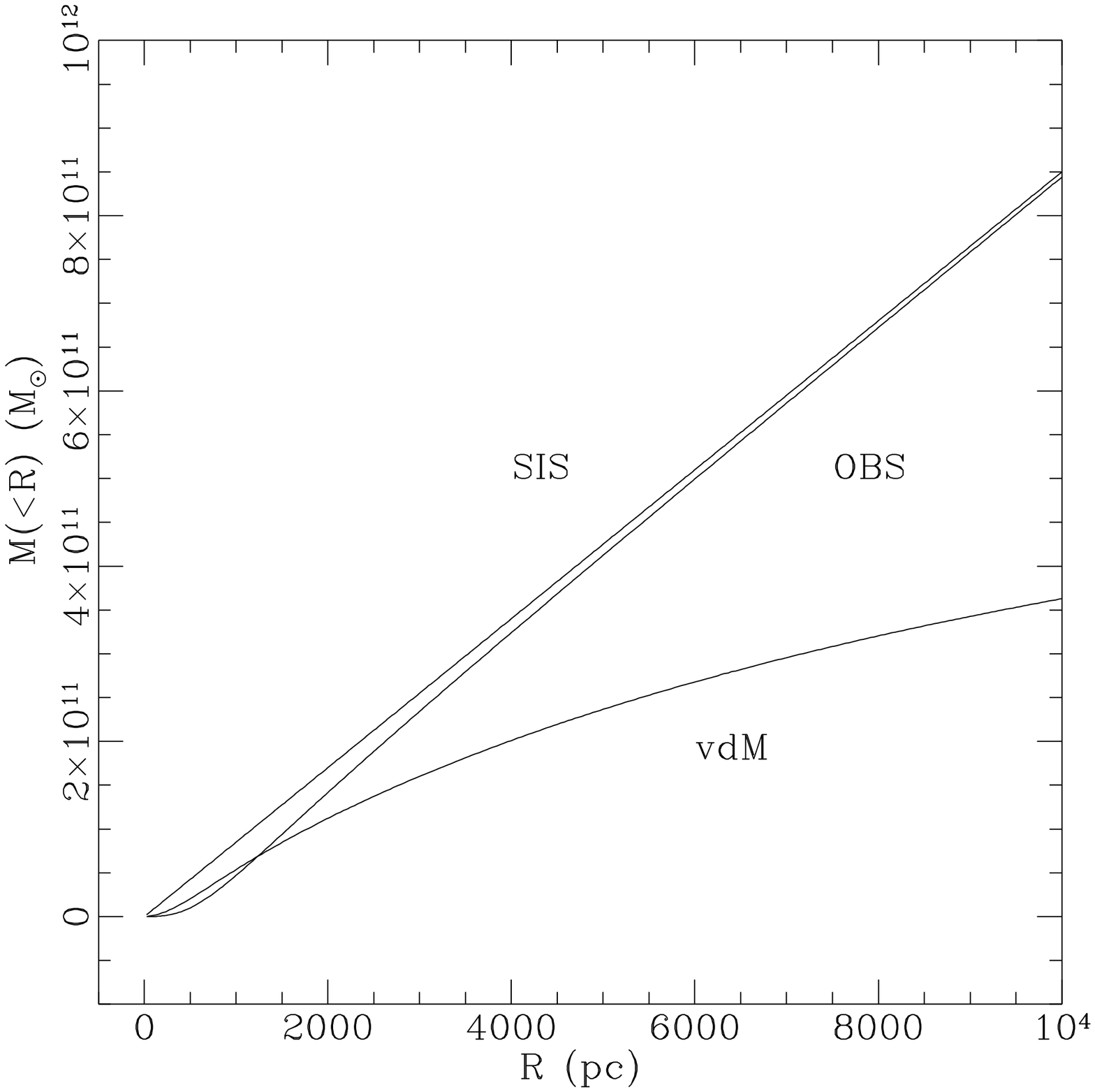}
\caption{Comparison of singular isothermal sphere (SIS), softened isothermal 
		sphere used by OBS, and the luminosity profile
		determined by van der Marel (1994).  The isothermal
		models agree with the observed profile at
		small radii and continue to rise linearly with the
		dark matter halo.
}
\label{fig:mass_comp}
\end{figure}

	 We adopt a distance of 16 Mpc to M87 which corresponds to
$H_0=81.5\kms\mpc^{-1}$ (van der Marel 1994).  This defines a length
scale of 77.6 pc per second of arc.  Lauer \& Kormendy (1986) adopted
$H_0=75\kms\mpc^{-1}$, giving a distance of 17.4 Mpc and defining a length 
scale of 84.1 pc per second of arc.  The more recent estimate by
Elson \& Santiago (1996) based on the Cepheid calibration of Freedman
et al. (1994) gives a length scale of 87 pc per second of arc which is
closer to the latter estimate but not strongly discrepant from the
first.  OBS apparently adopted the length scale of 132 pc per second
of arc, corresponding to $H_0=50\kms\mpc^{-1}$.

	For the singular isothermal sphere (SIS), $\Phi(r)=v_0^2\ln
r$, we adopt $v_0=606\kms$ (OBS).  This defines a velocity dispersion
of $\sigma=350\kms$ for an initial distribution of luminous matter
with $\eta^{-1}={1\over 3}$.  Figure \ref{fig:mass_comp} compares the
mass of the SIS with the softened isothermal sphere used by OBS, and
van der Marel's (1994) estimate derived using the luminosity profile
and dynamical modeling.

	The results presented here do not significantly depend on
choice of singular or softened potential.  The 10\% mass difference at
3 kpc leads to 5\% faster evolution in the SIS since the dynamical
time scale goes as $\rho^{-1/2}$ in tidally-limited clusters.  This is
well within the 4.7 kpc core radius of the cluster system determined
by McLaughlin (1995).  Conversely, the tidal field in the the slowly
rising core region of the softened model is closer to that in a
Keplerian potential which is stronger and so balances the larger mass
of the SIS.

\section{Maximum likelihood estimation of model parameters}
\label{sec:mlAp}

	A joint $\chi^2$-maximum likelihood estimator is used in
\S\ref{sec:results} to fit the dynamical models to V-band photometric
data and binned surface density data of the M87 cluster system (MHH).
THe expected surface density profiles are the sum of the dynamically
evolved model surface density, $S(r,v;\theta)$, derived from
distributions in \S\ref{sec:evol}, and background surface density,
$\sigma_0(V)$, multiplied by the incompleteness factor, $f(x,y,V)$
which represents the probability of detecting a cluster of given
magnitude at a particular location in the field.  The likelihood
statistic
\begin{equation}
L=\prod_j f(x_j,y_j,V_j){\bigl[} S(r_j,V_j;\theta)+\sigma_0(V_j){\bigr ]},
\end{equation}
defines the posterior probability of the data given the model.  The
$\chi^2$ statistic
\begin{equation}
P_{\chi^2}=\prod_i {1\over \sqrt{2\pi\sigma_i}}e^{-{1\over 2}
	{\bigl(}{S_i-S(r,\theta)\over\sigma_i}{\bigr)}^2}.
\end{equation}
defines the posterior probability of the binned data given the model.
The total posterior probability is then
\begin{equation}
P=P_{\chi^2}\times L
\end{equation}
and the parameters which maximize the posterior probability of the
data are the best estimates. 


\begin{thebibliography}{}
\bibitem[]{}
Binney, J. \& Tremaine, S. 1987, Galactic Dynamics, (Princeton: 
	Princeton U. Press)
\bibitem[]{}
Blakeslee, J. P. \& Tonry, J. L. 1995, ApJ, 442, 579
\bibitem[]{}
Brodie, J. 1993, in The Globular Cluster-Galaxy Connection, ASP
	Conference Series, v. 48, ed G. Smith \& J. Brodie, p. 483
\bibitem[]{}
Chernoff, D. \& Weinberg, M.~D. 1990, ApJ, 351, 121
\bibitem[]{}
Cohn, H. 1979, ApJ, 234, 1036
\bibitem[]{}
Djorgovski, S. D. \& Davis, M. 1987, ApJ, 313, 59
\bibitem[]{}
Elson, R. \& Santiago, B. 1996, MNRAS, 278, 617
\bibitem[]{}
Faber, S. M. 1995, private communication
\bibitem[]{}
Faber, S. M. \& Jackson, R. E. 1976, ApJ, 204, 668
\bibitem[]{}
Faber, S. M., Dressler, A., Davies, R. L., Burstein, D., Lynden-Bell, D.,
	Terlevich, R., Wegner, G. 1987, in Nearly Normal Galaxies: From the
	Planck Time to the Present, ed. S. M. Faber, 
	(New York: Springer-Verlag), p. 175
\bibitem[]{}
Freedman, W. L., Madore, B. F., Mould, J. R., Hill, R., Ferrarese, L.,
	Kenicutt, R. C., Saha, A., Stetson, P. B., Graham, J. A., Ford, H.,
	Hoessel, J. G., Huchra, J., Hughes, S. M., Illingworth, G. D. 1994, 
	Nature, 371, 757
\bibitem[]{}
Harris, W. E. 1991, ARAA, 29, 543
\bibitem[]{}
Harris, W. E. 1996, private communication
\bibitem[]{}
Harris, W. E. \& Pudritz, R. E. 1994, ApJ, 429, 177
\bibitem[]{}
Huchra, J. 1988, in The Harlow-Shapley Symposium on Globular Clusters in
	Galaxies, ed. J. Grindlay \& A. Philip, (Boston: Kluwer),
	p. 255
\bibitem[]{}
Kissler, M., Richtler, T., Held, E. V., Grebel, E. K., Wagner, S. J., 
	Capaccioli, M. 1994, A\&A, 287, 463
\bibitem[]{}
Lee, H., Fahlman, G., \& Richer, H. 1991, ApJ, 366, 455
\bibitem[]{}
Lee, M. \& Geisler, D. 1993, in The Globular Cluster-Galaxy Connection, 
	ASP Conference Series, v. 48, ed G. Smith \& J. Brodie, p. 576
\bibitem[]{}
Lauer, T. R. 1988, ApJ, 325, 49
\bibitem[]{}
Lauer, T. R. \& Kormendy, J. 1986, ApJ, 303, L1
\bibitem[]{}
Martin, B. 1971, Statistics for Physicists, (New York: Academic 
	Press)
\bibitem[]{}
Mihos, J. C. \& Hernquist, L. 1994, ApJ, 437, L47
\bibitem[]{}
Murali, C. \& Weinberg, M. D. 1996, MNRAS, submitted (MW)
\bibitem[]{}
McLaughlin, D. E. 1995, AJ, 109, 2034
\bibitem[]{}
McLaughlin, D. E. \& Harris, W. E. 1995, private communication
\bibitem[]{}
McLaughlin, D. E., Harris, W. E., \& Hanes, D. A. 1994, ApJ, 422, 486 (MHH)
\bibitem[]{}
McLaughlin, D. E. \& Pudritz, R. E. 1996, ApJ, 457, 578
\bibitem[]{}
Ostriker, J., Binney, J. \& Saha, P. 1989, MNRAS, 241, 849 (OBS)
\bibitem[]{}
Ostriker, J. \& Tremaine, S. 1975, ApJ, 202, L113
\bibitem[]{}
Pahre, M. A., Djorgovski, S. G. and de Carvalho, R. R. 1995,            
ApJ, 453, L17
\bibitem[]{}
Santiago, B. X \& Djorgovski, S. G. 1993, MNRAS, 261, 753
\bibitem[]{}
Strom, S., Strom, K., Wells, D., Forte, J., Smith., M. \& Harris, W.
	1981, ApJ, 245
\bibitem[]{}
van den Bergh, S. 1995, preprint
\bibitem[]{}
van der Marel, R., 1994 MNRAS, 270, 271
\bibitem[]{}
Weinberg, M. D. 1994, AJ, 108, 1414
\bibitem[]{}
Zepf, S. E., \& Ashman, K. M. 1993, MNRAS, 264, 611
\bibitem[]{}
Zepf, S. E., Geisler, D. \& Ashman, K. M. 1994, ApJ, 435, L117

\end{thebibliography}
\end{document}